\def\kB{k_{\text B}}

\def\sgn{{\text{sgn\,}}}
\def\be{\begin{equation}}
\def\ee{\end{equation}}
\def\bea{\begin{eqnarray}}
\def\eea{\end{eqnarray}}
\def\bse{\begin{subequations}}
\def\ese{\end{subequations}}

\def\erf{{\text{erf}}\,}

\documentclass[prb,twocolumn,amsmath,amssymb,eqsecnum,preprintnumbers]{revtex4}
\usepackage{graphicx}
\usepackage{dcolumn}
\usepackage{bm}
\usepackage{bbm}
\usepackage{color}
\usepackage{enumitem}
\begin{document}

\title{Rigidity and Superfast Signal Propagation in Fluids and Solids in Non-Equilibrium Steady States}

\author{T.R. Kirkpatrick$^{1}$, D. Belitz$^{2,3}$ and J.R. Dorfman$^{1}$}

\affiliation{$^{1}$Institute for Physical Science and Technology, University of Maryland, College Park, MD 20742, USA\\
 $^{2}$Department of Physics and Institute for Fundamental Science, University of Oregon, Eugene, OR 97403, USA \\
 $^{3}$ Materials Science Institute, University of Oregon, Eugene, OR 97403, USA}

\date{\today}
\begin{abstract}

In the 1980s it was theoretically predicted that correlations of various observables in a fluid in a non-equilibrium steady state (NESS) are 
extraordinarily long-ranged, extending, in a well-defined sense, over the size of the system. This is to be contrasted with correlations in an 
equilibrium fluid, whose range is typically just a few particle diameters. These NESS correlations were later confirmed by numerous
experimental studies. Unlike long-ranged correlations at critical points, these correlations are generic in the sense that they exist for any 
temperature as long as the system is in a NESS. In equilibrium systems, generic long-ranged correlations are caused by spontaneously 
broken continuous symmetries and are associated with a generalized rigidity, which in turn leads to a new propagating excitation or mode. 
For example, in a solid, spatial rigidity leads to transverse sound waves, while in a superfluid, phase rigidity leads to temperature waves 
known as second sound at finite temperatures, and phonons at zero temperature. More generally, long-ranged spatial correlations imply 
rigidity irrespective of their physical origin. This implies that a fluid in a NESS should also display a type of rigidity and related anomalous 
transport behavior. Here we show that this is indeed the case. For the particular case of a simple fluid in a constant temperature gradient, 
the anomalous transport behavior takes the form of a super-diffusive spread of a constant-pressure temperature perturbation. We also 
discuss the case of an elastic solid, where we predict a spread that is faster than ballistic.
\end{abstract}

\maketitle

\section{Introduction}
\label{sec:I}

Correlations in a fluid in equilibrium are very short-ranged on a macroscopic scale. For instance, for distances large compared to a 
molecular diameter the temperature-temperature correlation function (TTCF) is given by \cite{Landau_Lifshitz_V_1980}
\bse
\label{eqs:1.1}
\be
\langle\delta{T}({\bm r})\delta{T}({\bm r}')\rangle = \frac{\kB T_{\text{eq}}^2}{c_V}\,\delta({\bm r} - {\bm r}')\ ,
\label{eq:1.1a}
\ee
or, in wave-number space,
\be
\langle\vert\delta T({\bm k})\vert^2\rangle = \frac{\kB T_{\text{eq}}^2}{c_V}\ .
\label{eq:1.1b}
\ee
\ese
Here $\delta{T({\bm r}})=T({\bm r})-T_{\text{eq}}$ is the temperature fluctuation, with $T({\bm r})$ the local fluctuating temperature and 
$T_{\text{eq}}$ the equilibrium temperature, $\kB$ is Boltzmann's constant, $c_V$ is the specific heat per volume at constant 
volume, and the angular brackets denote an average over an equilibrium ensemble.

\begin{figure}[b]
\includegraphics[width=6cm]{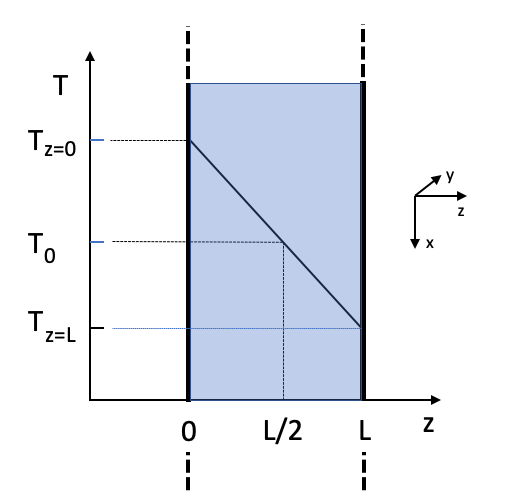}
\caption{A fluid with a linear temperature profile between two parallel confining plates.}
\label{fig:1}
\end{figure}
In a non-equilibrium steady state (NESS), by contrast, spatial correlations behave dramatically differently. To be specific, consider a simple fluid 
subject to a constant temperature gradient in the $z$-direction, as illustrated in Fig.~\ref{fig:1}. We consider a system that has a spatial extent $L$ in the 
$z$-direction and is infinite in the directions perpendicular to the direction of the gradient. For this case, the TTCF in wave-number space 
is\cite{Kirkpatrick_Cohen_Dorfman_1982c, Ortiz_Sengers_2007, Dorfman_Kirkpatrick_Sengers_1994}
\be
\langle|\delta{T({\bm k}})|^2\rangle=\frac{\kB T_0^2}{c_V} + (\partial_z T)^2\,\frac{\kB T_0}{\rho D_T(\nu+D_T)}\frac{{\hat{\bm k}}_{\perp}^2}{k^4}\ .
\label{eq:1.2}
\ee
Here the angular brackets denote a non-equilibrium (NE) ensemble, and only the leading small-$k$ (large distance) terms have been retained 
for the NE contribution. In Eq.~(\ref{eq:1.2}), $\rho$ is the mass density, $D_T$ is the thermal diffusivity, and $\nu$ is the kinematic viscosity. 
$T_0$ is the spatially averaged temperature of the NE fluid, and $\partial_z T = \text{const}$ is the constant temperature gradient.
$\hat{\bm k} = {\bm k}/k$ with $k = \vert{\bm k}\vert$ is the unit wave vector, and $\hat{\bm k}_{\perp}$ is its component 
perpendicular to the direction of the temperature gradient, i.e., $\hat{\bm k}_{\perp} = (k_x,k_y)/k$. All thermophysical quantities in 
Eq.~(\ref{eq:1.2}) should be interpreted as spatially averaged. 
Slip boundary conditions have been used, which leads to $k_z=N\pi/L$ with $N$ a positive integer. The $k^{-4}$ small-$k$ singularity in the 
NE term in Eq.~(\ref{eq:1.2}) indicates that very long-ranged correlations result from the temperature gradient. This result has been derived by kinetic 
theory,\cite{Kirkpatrick_Cohen_Dorfman_1982a, Kirkpatrick_Cohen_Dorfman_1982c} 
mode-coupling theory,\cite{Kirkpatrick_Cohen_Dorfman_1982a, Kirkpatrick_Cohen_Dorfman_1982c} 
and fluctuating hydrodynamics.\cite{Ronis_Procaccia_1982, Ortiz_Sengers_2007} The general equivalence of kinetic theory and 
fluctuating hydrodynamics for computing long-ranged correlations in a NESS was shown in Ref.~\onlinecite{Kirkpatrick_Cohen_Dorfman_1982a}.

The real-space TTCF can be written
\bea
\langle\delta{T}({\bm r})\delta{T}({\bm r}')\rangle &=& \frac{\kB T_0^2}{c_V}\delta({\bm r} - {\bm r}') \qquad\qquad
\nonumber\\
&& \hskip -25pt +\frac{\kB T_0}{\rho D_T(\nu+D_T)}\,G_{\text{NE}}(r_{\perp}, z, z')\ ,
\label{eq:1.3}
\eea
where $r_{\perp} = \sqrt{(x-x')^2 + (y-y')^2}$. 
In general, $G_{\text{NE}}$ is a complicated function of its arguments. Two simple limits are (see Sec.~7.5 in Ref.~\onlinecite{Ortiz_Sengers_2007})
\bse
\label{eqs:1.4}
\be
G_{\text{NE}}(r_{\perp}=0, z\ll L/2, z'\ll L/2)=\frac{L}{16}\left[1-2\frac{|z-z'|}{L}\right]
\label{eq:1.4a}
\ee
and
\be
G_{\text{NE}}(r_{\perp}\ll L, z=0, z'=0)=\frac{L}{16}\left[1-3\frac{r_\perp}{L}\right] \ .
\label{eq:1.4b}
\ee
\ese
Note that in both limits the real-space correlations scale with the system size $L$, and the correlations decay on the same scale.
For instance, if the system size $L$ is scaled by a factor of $b>1$, and the distance $\vert z - z'\vert$ in Eq.~(\ref{eq:1.4a}) is scaled by
the same factor, then the correlation $G_{\text{NE}}$ increases by a factor of $b$. However, if $\vert z - z'\vert$ increases at fixed $L$, then
 $G_{\text{NE}}$ decreases. This is the real-space manifestation of the $1/k^4$ singularity in Eq.~(\ref{eq:1.2}).

Physically, one expects these long-ranged correlations to have dynamical consequences in a macroscopic description of a fluid, independent 
of thermal fluctuation effects. That is, if the fluid is so strongly correlated that spatial correlations extend throughout the entire system, 
which represents a generalized rigidity,
then a perturbation at one point in the fluid should propagate infinitely faster, in a scaling sense, then a diffusive process. Indeed, we will
show that a temperature perturbation at $t=0$ at a distance $R$ from the observer is detectable at a time $t = R/v_0$, with $v_0$ a
characteristic velocity, rather than at the much longer diffusive time scale $t = R^2/D_T$ that characterizes an equilibrium fluid.
This is analogous to what happens in equilibrium systems if a spontaneously broken continuous symmetry
leads to generic long-ranged correlations\cite{Forster_1975, Anderson_1984} that endow the system with a generalized
rigidity property.\cite{Anderson_1984, rigidity_footnote} In that case the broken symmetry leads to Goldstone modes, which typically are
propagating and emerge in addition to any soft modes that may be present in the absence of symmetry breaking.
For example, in a solid the long-ranged spatial correlations (namely, displacement fluctuations that scale as $1/k^2$),
and the associated rigidity (represented by a nonzero shear modulus) lead 
to transverse sound waves.\cite{Chaikin_Lubensky_1995, Martin_Parodi_Pershan_1972} That is, in a fluid in equilibrium the transverse 
modes are diffusive, while in a solid they are propagating. Another example is second sound in superfluids, where long-ranged phase 
correlations couple to energy-density fluctuations to form a propagating mode at nonzero temperature, second sound, that is 
a constant-pressure temperature wave.\cite{Chaikin_Lubensky_1995} In normal fluids, by contrast, temperature perturbations at 
constant pressure are diffusive.\cite{Forster_1975} At zero temperature these second-sound excitations are phonons with the same 
linear dispersion relation and the same speed of (second) sound as the Goldstone mode, namely, the single-particle Bogoliubov 
excitations.\cite{Gavoret_Nozieres_1964} Yet another example is the magnon in the magnetically ordered phase of a Heisenberg 
ferromagnet, which is a propagating spin wave, whereas in the paramagnetic phase the corresponding transverse spin modes are 
diffusive.\cite{Forster_1975, Chaikin_Lubensky_1995} In all of these examples, the long-ranged static correlations,
and the associated rigidity, lead to a signal propagation that scales linearly with time, as opposed to a diffusive process, where it
scales as the square root of time.

It is the purpose of the present paper to examine the dynamic consequences of the generic long-ranged correlations in a fluid in a 
NESS. We will show that, in a well-defined sense, temperature fluctuations in a fluid in a NESS spread as fast as a signal transmitted 
by a propagating wave, and in a solid they spread even faster.

An outline of this paper is as follows. In Section~\ref{sec:II} we give the fundamental equations describing both fluctuations in NESS and 
macroscopic perturbations about a NESS. We then summarize the results for dynamical fluctuations in a NESS, and derive an
effective equation for temperature fluctuations that sheds light on the structure of these results. In Section~\ref{sec:III} 
we discuss rigidity in a NESS realized by a constant temperature gradient, and show that it leads to signal propagation that is
super-diffusive in a fluid, and faster than ballistic in a solid. In Section~\ref{sec:IV} we conclude with a discussion of our results.

\section{Langevin equations, and the temperature-temperature time correlation function in a NESS}
\label{sec:II}

In this section we give the Langevin equations that describe temperature and velocity fluctuations about a NESS. These equations are then 
used to obtain the dynamic fluctuation about a NESS, and Eq.~(\ref{eq:1.2}). Both of these quantities can be directly measured by light scattering 
experiments. We also present an alternative procedure that derives an effective Langevin equation for temperature fluctuations only. 

\subsection{Langevin equations}
\label{subsec:II.A}

Ignoring fast sound-mode or pressure-fluctuation effects, the Langevin equations describing fluctuations in a simple fluid in a thermal gradient, 
see Fig.1, are\cite{Landau_Lifshitz_VI_1987, Ortiz_Sengers_2007}
%
%
\bse
\label{eqs:2.1}
\be
\partial_t\delta T({\bm r}, t) + v_z({\bm r}, t)\partial_z T = D_T\nabla^2\delta T({\bm r}, t) + Q({\bm r}, t)
\label{eq:2.1a}
\ee
and,
\be
\partial_t v_z({\bm r}, t) = \nu\nabla^2 v_z({\bm r}, t)+P_z({\bm r}, t)
\label{eq:2.1b}
\ee
\ese
Here $v_z$ is the $z$-component of the fluctuating transverse velocity,\cite{transverse_footnote}
$D_T$ is the thermal diffusivity, and $\nu$ is the kinematic viscosity. $Q$ and $P_z$ are Langevin forces that are 
Gaussian distributed and delta-correlated in space and time,
\bse
\label{eqs:2.2}
\bea
\langle Q({\bm r},t)Q({\bm r}',t')\rangle &=& \frac{2\kB T_0^2}{c_p}D_Tk^2 \delta({\bm r}-{\bm r}')\delta(t-t')
\nonumber\\
&\equiv& G_{QQ}({\bm r},t;{\bm r}',t')\ ,
\label{eq:2.2a}\\
\langle P_z({\bm r},t)P_z({\bm r}',t')\rangle &=& \frac{2\kB T_0}{\rho}\nu {\bm k}_{\perp}^2 \delta({\bm r}-{\bm r}')\delta(t-t')
\nonumber\\
&\equiv& G_{PP}({\bm r},t;{\bm r}',t')\ ,
\label{eq:2.2b}
\eea
\ese
or, in wave-number space,
\bse
\label{eqs:2.3}
\bea
\langle Q({\bm k},t)Q({\bm k}',t')\rangle &=& \frac{2\kB T_0^2}{c_p}D_Tk^2 
                                                                                                                        \delta_{{\bm k},-{\bm k}'}\ \delta(t-t'), \quad
\label{eq:2.3a}\\
\langle P_z({\bm k},t)P_z({\bm k}',t')\rangle &=& \frac{2\kB T_0}{\rho}\nu {\bm k}_{\perp}^2 
                                                                                                                                               \delta_{{\bm k},-{\bm k}'}\ \delta(t-t')\ .\qquad\quad
\label{eq:2.3b}
\eea
\ese
Here $c_p$ is the specific heat at constant pressure. The prefactors on the right-hand sides of these equations reflect
the equilibrium correlations of the temperature at constant pressure, $\langle\vert\delta T({\bm k})\vert^2\rangle = \kB T_0^2/c_p$,
and of the velocity, $\langle\vert v_z({\bm k})\vert^2\rangle = \kB T_0/\rho$.\cite{Landau_Lifshitz_V_1980} The cross correlations 
$\langle Q P\rangle =0$ vanish since there is no kinetic coefficient that couples $\delta T$ and $v_z$. For the validity of these 
equations in the context of long-range correlations in a NESS, see the discussion in Sec.~\ref{sec:IV}.

\subsection{The TTCF}
\label{subsec:II.B}

Solving these equations for the TTCF by Fourier transforming in space and time and then transforming back to time 
gives\cite{Kirkpatrick_Cohen_Dorfman_1982c, Ortiz_Sengers_2007}
\bse
\label{eqs:2.4}
\bea
\langle\delta T({\bm k},t)\delta T^*({\bm k},0)\rangle &=& \frac{\kB T_0^2}{c_p}\Bigl[(1+A_T(k))\exp(-D_Tk^2|t|)
\nonumber\\
&& \hskip 20pt - A_{\nu}(k)\exp(-\nu k^2|t|)\Bigr]\ ,
\label{eq:2.4a}
\eea
where
\be
A_{\nu}(k) = \frac{D_T}{\nu}A_T(k) = \frac{c_p}{T_0}\frac{{\hat{\bm k}_{\perp}}^2(\partial_z T)^2}{(\nu^2-D_T^2)k^4}\ .
\label{eq:2.4b}
\ee
\ese
Note that $A_T$ and $A_{\nu}$ are singular for $k \to 0$ and scale as $1/k^4$. Setting $t=0$ in Eq.~(\ref{eq:2.4a}) gives Eq.~(\ref{eq:1.2})
adapted for the case of constant pressure.\cite{pressure_fluctuations_footnote}

Equation~(\ref{eq:2.4a}) can be directly measured in small angle light scattering.\cite{Kirkpatrick_Cohen_Dorfman_1982c, Ortiz_Sengers_2007} 
The results are shown in Fig.~2. There are no adjustable parameters in the fit, all thermo-physical properties are taken from other experimental data.  
Note how large the effect is: For the wave numbers and temperature gradients in the experiment, the NE contribution is much larger than the
equilibrium one. The conclusion is that the long-ranged correlations in a NESS are well confirmed by these experiments as well as by many 
others.\cite{more_experiments_footnote}
\begin{figure}[t]
\includegraphics[width=6cm]{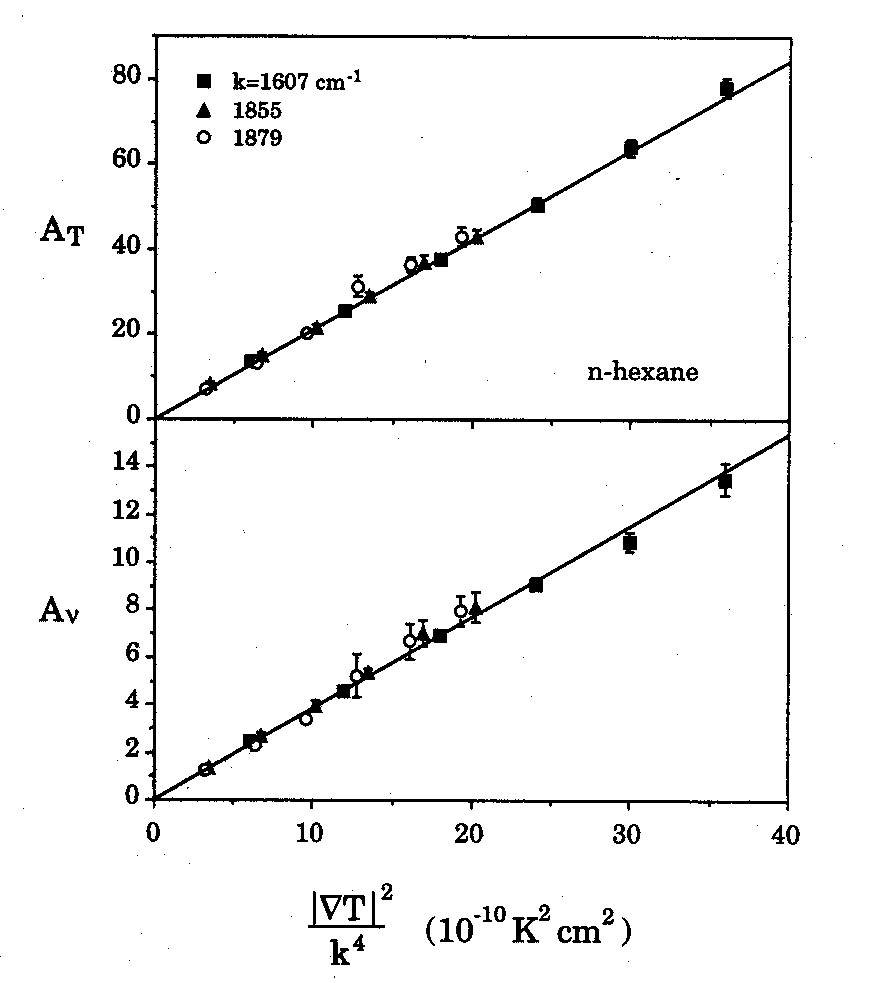}
\caption{The amplitudes $A_T$ and $A_{\nu}$, Eq.~(\ref{eq:2.4b}), measured in liquid hexane at $25^o\,{\rm C}$ as a function of 
              $(\partial_z T)^2/{\bf k}^4$. The symbols indicate experimental data for three different wave numbers. The solid lines represent 
              the values predicted by Eq.~(\ref{eq:2.4b}). From Ref.~\onlinecite{Li_et_al_1994}.}
\label{fig:2}
\end{figure}

An interesting aspect of Eqs.~(\ref{eqs:2.4}) is the fact that the time dependence is entirely diffusive: A Laplace transform of either of the two
terms in Eq.~(\ref{eq:2.4a}) has the form of an ordinary diffusion pole
\bse
\label{eqs:2.5}
\be
\mathcal{D}(k,z) = \frac{A(k)}{z + iDk^2}
\label{eq:2.5a}
\ee
with a spectrum
\be
\mathcal{D}''(k,\omega) = \text{Im}\,\mathcal{D}(k,\omega + i0) = A(k)\,\frac{D k^2}{\omega^2 + D^2 k^4}\ .
\label{eq:2.5b}
\ee
\ese
Here $z$ is a complex frequency with $\text{Im}\,z > 0$ and $D$ is a diffusivity that in the present context can be either $D_T$ or $\nu$. 
What is anomalous is the prefactor $A(k)$, which represents a static susceptibility
\be
\int_{-\infty}^{\infty} \frac{d\omega}{\pi}\,\mathcal{D}''(k,\omega) = A(k) \sim 1/k^4
\label{eq:2.6}
\ee
that is highly singular in the limit $k\to 0$, scaling as $1/k^4$, as a result of the non-equilibrium effects, see Eq.~(\ref{eq:2.4b}). The diffusivity 
is related to a generalized conductivity $\sigma$ via an Einstein relation $\sigma = D\chi$, with $\chi$ another static susceptibility that is not 
qualitatively affected by the nonequilibrium fluctuations. It is illustrative to consider this structure from another angle by deriving an effective 
equation for the temperature fluctuations only, as we do in the following subsection.

\subsection{Effective Langevin equation for temperature fluctuations}
\label{subsec:II.C}

Let us rewrite the Langevin equations (\ref{eqs:2.1}, \ref{eqs:2.2}), using a Martin-Siggia-Rose formalism.\cite{Martin_Siggia_Rose_1973,
Bausch_Janssen_Wagner_1976, DeDominicis_Peliti_1978} The starting point is the stochastic `partition function'
\bea
Z[Q,P]\!\! &=& \!\!\int D[\delta T, v_z]\ \delta\!\left[\partial_t \delta T + v_z \partial_z T - D_T\nabla^2 \delta T - Q\right]
\nonumber\\
&& \times\delta\!\left[\partial_t v_z - \nu\nabla^2 v_z - P\right]\,J
 \label{eq:2.7}
 \eea
Here the integrations and the $\delta$-functions are to be understood in a functional sense, and $J$ is a Jacobian associated with
the arguments of the $\delta$-functions that ensures that $Z[Q,P] = 1$. By adding sources for $\delta T$ and $v_z$ one can turn $Z$
into a generating functional for correlation functions. This will not be important for what follows, and neither will the Jacobian, which for 
our linearized theory is independent of the fields.\cite{Jacobian_footnote} In what follows we will ignore the Jacobian, as well as
constant prefactors that arise from Gaussian integrals. The next step is to introduce auxiliary `conjugate' fields
$\widetilde{\delta T}$ and $\tilde{v}_z$ to rewrite the functional $\delta$-functions in terms of auxiliary integrals:
\bse
\label{eqs:2.8}
\bea
Z[Q,P] &=& \int D[\delta T, v_z, \widetilde{\delta T}, \tilde{v}_z]\ 
\nonumber\\
&& \times e^{i \left(\widetilde{\delta T} \bigl\vert \partial_t \delta T + v_z \partial_z T - D_T\nabla^2 \delta T - Q\right)}
\nonumber\\
&& \times e^{i \left( \tilde{v}_z \bigl\vert \partial_t v_z - \nu\nabla^2 v_z - P\right)}\ ,
\label{eq:2.8a}
\eea
where we have defined a scalar product
\bea
\left( A\vert B\right) &=& \int d{\bm r} dt\,A({\bm r},t)\,B({\bm r},t) 
\nonumber\\
&=& \frac{1}{V}\sum_{\bm k} \int d\omega\,A({\bm k},\omega)\,B(-{\bm k}-\omega)\ .\quad
\label{eq:2.8b}
\eea
\ese
We next integrate out the Langevin forces $Q$ and $P$,
using Gaussian distributions with second moments given by Eqs.~(\ref{eqs:2.2}, \ref{eqs:2.3}). It is most convenient to work in Fourier space,
which makes the second moments\cite{frequency_space_footnote}
\bse
\label{eqs:2.9}
\bea
\langle \vert Q({\bm k},\omega)\vert^2\rangle &\equiv& G_{QQ}({\bm k},\omega) = \frac{2\kB T_0^2}{c_p}\,D_T k^2\ ,\quad
\label{eq:2.9a}\\
\langle \vert P_z({\bm k},\omega)\vert^2\rangle &\equiv& G_{PP}({\bm k},\omega) = \frac{2\kB T_0}{\rho}\,\nu{\bm k}_{\perp}^2\ .
\label{eq:2.9b}
\eea
\ese
The moments are frequency independent due to the delta-correlations in time space.
We obtain
\bea
Z &=& \int D[Q,P]\,Z[Q,P]\,e^{-\frac{1}{2}\left(Q \bigl\vert G^{-1}_{QQ}\bigl\vert Q\right) -\frac{1}{2}\left(P_z \bigl\vert G^{-1}_{PP}\bigl\vert P_z\right)}
\nonumber\\
   &=& \int D[\delta T, v_z, \widetilde{\delta T}, \tilde{v}_z]\, e^{i \left(\widetilde{\delta T} \bigl\vert (-i\omega + D_T{\bm k}^2) \delta T + (\partial_z T) v_z\right)}
\nonumber\\
&& \times  e^{i \left( \tilde{v}_z \bigl\vert (-i\omega + \nu{\bm k}^2) v_z \right)}\,e^{-\frac{1}{2}\left(\widetilde{\delta T}\bigl\vert G_{QQ}
     \bigl\vert\widetilde{\delta T}\right)  - \frac{1}{2}\left(\tilde{v}_z\bigl\vert G_{PP} \bigl\vert\tilde{v}_z\right) }
\nonumber\\
\label{eq:2.10}
\eea 
We now integrate out $\tilde{v}_z$, which produces a term that is quadratic in $v_z$, and finally we integrate out $v_z$. This procedure yields
\bea
Z &=& \int D[\delta T, \widetilde{\delta T}]\,e^{i\left(\widetilde{\delta T}\bigl\vert(-i\omega + D_T{\bm k}^2)\delta T\right)}
\nonumber\\
&& \times e^{- \left( \widetilde{\delta T}\left[\frac{\kB T_0^2}{c_p} D_T {\bm k}^2 + \frac{(\partial_z T)^2 \kB T_0 \nu{\bm k}_{\perp}^2}{\rho(\omega^2 + \nu^2{\bm k}^2)}\right]  \widetilde{\delta T}\right)}
\label{eq:2.11}
\eea
A comparison with Eq.~(\ref{eq:2.10}) shows that this result is equivalent to a fluctuating diffusion equation for $\delta T$ only,
\bse
\label{eqs:2.12}
\be
\partial_t \delta T = D_T \nabla^2 \delta T + Q_r({\bm r},t)\ ,
\label{eq:2.12a}
\ee
with a renormalized fluctuating force $Q_r$ that is Gaussian distributed with a second moment
\be
\langle\vert Q_r({\bm k},\omega)\vert^2\rangle = \frac{2\kB T_0^2}{c_p}\,D_T k^2 + \frac{2\kB T_0}{\rho}\,
   \frac{(\partial_z T)^2 \nu {\bm k}_{\perp}^2}{\omega^2 + \nu^2 k^4}\ .
\label{eq:2.12b}   
\ee
\ese
The salient point is that the fluctuating force gets renormalized, and becomes long ranged. Indeed, the relative scaling of the
non-equilibrium term compared to the equilibrium one is $1/k^4$, just as in the TTCF, Eqs.~(\ref{eqs:2.4}). The diffusion
coefficient, on the other hand, does not get renormalized, 
in agreement with the discussion in Sec.~\ref{subsec:II.B}. However, since we have integrated out the velocity fluctuations, this 
description of effective anomalous temperature diffusion does not reflect the anomalous behavior caused by an initial velocity 
perturbation discussed in the next section (see Eq.~(\ref{eq:3.2}) below).

It is important to note that in this effective description of temperature fluctuations an assumption of a delta-correlated fluctuating
force would be incorrect: the coupling to the, now hidden, velocity degrees of freedom leads to an emergent long-rangedness
of the fluctuating force. See also Point 3. in Sec.~\ref{sec:IV}.

\section{A Dynamical Consequence of Rigidity in a NESS}
\label{sec:III}

In this section we ignore thermal fluctuation effects and simply consider how far a macroscopic perturbation  at one point in the fluid travels in a time $t$. 
The equations in Sec.~\ref{sec:II} are diffusive, so in equilibrium a macroscopic perturbation diffuses a distance proportional to $t^{1/2}$. As we will show,
in a fluid in a NESS this distance scales as $t$, which is the same result as for a propagating perturbation. We will then show, by analogous
arguments, that the same effect in a solid leads to the distance scaling as $t^{3/2}$, i.e., information about the perturbation travels faster than
ballistically.

\subsection{Fluids}
\label{subsec:III.A}

The equations describing macroscopic perturbations about the NESS in a fluid are Eqs.~(\ref{eqs:2.1}) without the fluctuating forces:
\bse
\label{eqs:3.1}
\be
\partial_t\delta T({\bm r}, t) + v_z({\bm r}, t)\,\partial_z T = D_T\nabla^2\delta T({\bm r}, t)\ ,
\label{eq:3.1a}
\ee
and
\be
\partial_t v_z({\bm r}, t) = \nu\nabla^2 v_z({\bm r}, t)\ ,
\label{eq:3.1b}
\ee
\ese
where $\delta T$ and $v_z$ are macroscopic perturbations specified by initial conditions $\delta T({\bm r}, t=0) = \delta T^{(0)}({\bm r})$ and 
$v_z({\bm r}, t=0) = v_z^{(0)}({\bm r})$.

These equations are easy to solve for $\delta T$ by using a spatial Fourier transform and a temporal Laplace transform. Transforming
back to real time yields
\bea
\delta T({\bm k}, t) &=& \delta T^{(0)}({\bm k})\,e^{-D_T k^2 t } 
\nonumber\\
&& \hskip 0pt +\frac{v_z^{(0)}({\bm k})\,\partial_z T}{k^2(\nu - D_T)}\left[e^{-\nu k^2 t} - e^{-D_T k^2 t}\right]. \quad
\label{eq:3.2}
\eea
Note that the time dependence of all terms in Eq.~(\ref{eq:3.2}) is diffusive. However, since $k^2$ scales as $1/t$, the $1/k^2$ factor in the NE part of 
Eq.~(\ref{eq:3.2}) suggests that the spread of the initial temperature perturbation is effectively faster than diffusive. To make this precise, we assume 
strongly localized initial perturbations, which in our macroscopic description are represented by $\delta$-functions in space. Accordingly, we take 
\bse
\label{eqs:3.3}
\bea
\delta T^{(0)}({\bm k}) &=& \delta T^{(0)}
\label{eq:3.3a}\\
v_z^{(0)}({\bm k}) &=& v_z^{(0)}
\label{eq:3.3b}
\eea
\ese
to be indepen\-dent of the wave number. We can then perform a Fourier back transform into real space, which yields
\bse
\label{eqs:3.4}
\be
\delta T({\bm r},t) = \delta T_{\text{E}}(r,t) + \delta T_{\text{NE}}(r,t)\ ,
\label{eq:3.4a}
\ee
where $r = \vert{\bm r}\vert$. The equilibrium part has the usual diffusive form
\be
\delta T_{\text{E}}(r,t) = \frac{\delta T^{(0)}}{(4\pi D_T t)^{3/2}}\,e^{-r^2/4D_T t} 
\label{eq:3.4b}
\ee
For the non-equilibrium part one finds
\bea
\delta T_{\text{NE}}(r,t) &=& \frac{T_0\,\sgn((\partial_z T) v_z^{(0)})}{(\nu - D_T) t_0\, r}\,
\nonumber\\
&& \hskip -30pt \times \left[\erf\left(r/2\sqrt{\nu t}\right) - \erf\left(r/2\sqrt{D_T t}\right)\right]\ . \quad
\label{eq:3.4c}
\eea
Here $\erf$ is the error function, and 
\be
t_0 = 4\pi T_0/\vert(\partial_z T) v_z^{(0)}\vert
\label{eq:3.4d}
\ee
\ese
is a time scale that characterizes the NESS. Note that $\delta T_{\text{NE}}$ can be positive or negative; this has no physical significance.
We see that, for fixed $r/\sqrt{t}$, $\delta T_{\text{E}}$ scales as $1/t^{3/2}$, whereas $\delta T_{\text{NE}}$ scales
as $1/t^{1/2}$, consistent with Eq.~(\ref{eq:3.2}). As a result, their spatial moments have different time dependences. In particular,
\bea
\langle r^2\rangle &\equiv& \int d{\bm r}\ r^2\,\frac{\delta T_{\text{E}}({\bm r}, t)}{T_0} 
     +  \int d{\bm r}\ r^2\,\frac{\vert\delta T_{\text{NE}} ({\bm r}, t)\vert}{T_0}\nonumber\\
&=&  6 D_T\,\frac{\delta T^{(0)}}{T_0}\,t + \frac{\pi}{4}\, (\nu+D_T)\,\frac{t^2}{t_0}\ .
\label{eq:3.5}
\eea

Equation~(\ref{eq:3.5}) is our main result. The first term in this equation is the usual equilibrium result that the mean squared displacement grows 
linearly in time in a diffusive system. The second term,
\be
\langle r^2\rangle_{\text{NE}} = \frac{\pi}{4}\,(\nu+D_T)\,t^2/t_0 \ ,
\label{eq:3.6}
\ee
has the surprising property that it grows quadratically as a function of time, as is expected for a {\em propagating} mode.
In fact, writing it as 
\bse
\label{eqs:3.7}
\be
\langle r^2\rangle_{\text{NE}}= v_0^2\,t^2,
\label{eq:3.7a}
\ee
defines a characteristic velocity
\be
v_0 = \sqrt{\pi (\nu+D_T)/4 t_0}
\label{eq:3.7b}
\ee
\ese
that vanishes in the equilibrium limit where $t_0\to\infty$. 

As can be seen from the above derivation, this behavior, which is akin to ballistic propagation, is due to the non-equilibrium term proportional to 
$\partial_z T/k^2$ in Eq.~(\ref{eq:3.2}). In the fluctuation calculation of Sec.~\ref{sec:II} this term is effectively squared, which results in a term 
proportional to $(\partial_z T)^2/k^4$. The conclusion is that the long-ranged correlations in a NESS expressed by Eq.~(\ref{eq:1.2}) on one hand,
and the anomalous mean-squared spread of a perturbation expressed by Eqs.~(\ref{eq:3.6}, \ref{eqs:3.7}) on the other, have the same physical 
origin: They both are manifestations of rigidity in fluids in a NESS.

\subsection{Solids}
\label{subsec:III.B}

We now extend our discussion to the case of solids, 
which have rigidity even in equilibrium, as 
represented by a nonvanishing shear modulus. As we will see, the NE effects induced by a constant temperature gradient $\partial_z T$ lead
to an increased rigidity that leads to a TTCF that scales with the wave number as $(\partial_z T)^2/k^2$, and a mean square displacement that 
scales with time as $t^3$. That is, temperature perturbations in a NESS spread faster than ballistically.

For simplicity, we will consider an isotropic solid, and we focus on the coupling between temperature fluctuations and transverse
displacement fluctuations. The motivation for the latter is that in solids with a small shear modulus the transverse speed of sound
can be substantially less than the longitudinal one, meaning that the coupling is to a relatively soft, if still propagating, mode. The
applicable Langevin equations that replace Eqs.~(\ref{eqs:2.1}) now read
\bse
\label{eqs:3.8}
\bea
\partial_t \delta T({\bm r},t) &+& \partial_t u_z({\bm r},t)\partial_z T = D_T \nabla^2 \delta T({\bm r},t) + Q({\bm r}, t)\ ,
\nonumber\\
\label{eq:3.8a}\\
\partial_t^2 {\bm u}_{\perp}({\bm r},t) &=& c_{\perp}^2 \nabla^2 {\bm u}_{\perp}({\bm r},t) + \Gamma \nabla^2 \partial_t {\bm u}_{\perp}({\bm r},t) 
     + {\bm P}({\bm r},t)\ .
\nonumber\\
\label{eq:3.8b}
\eea
Here ${\bm u}_{\perp}$ is the transverse displacement field (i.e., $\partial_t {\bm u}_{\perp}$ is the transverse velocity), $u_z$ is the
$z$-component of ${\bm u}_{\perp}$, $c_{\perp}$ is the transverse sound velocity, and $\Gamma$ is the sound attenuation coefficient. 
The correlations of $Q$ are again given by Eqs.~(\ref{eq:2.2a}, \ref{eq:2.3a}), and those of ${\bm P}$ by
\bea
\langle P_i({\bm k},\omega) P_j(-{\bm k},-\omega)\rangle &=& 2 \delta_{ij}\langle\vert\partial_t u_i({\bm k})\vert^2\rangle\, {\bm k}_{\perp}^2 \Gamma 
\nonumber\\
&=& \delta_{ij} \frac{2\kB T_0}{\rho}\, {\bm k}_{\perp}^2 \Gamma\ .
\label{eq:3.8.c}
\eea
\ese
We specify initial conditions by
\bse
\label{eqs:3.9}
\bea
\delta T({\bm k},t=0) &=& \delta T^{(0)}\ ,
\label{eq:3.9a}\\
\left(\partial_t {\bm u}\right)_z({\bm k},t=0) &=& v_z^{(0)}\ ,
\eea
as in Eqs.~(\ref{eqs:3.3}), and
\be
u_z({\bm k},t=0) = 0\ .
\label{eq:3.9c}
\ee
\ese
The latter just represents our choice of the zero of time.

We now ignore the fluctuating forces and calculate the mean-squared displacement as we did for a fluid in
Sec.~\ref{subsec:III.A}. A spatial Fourier transform and a temporal Laplace transform yield
\bse
\label{eqs:3.10}
\bea
u_z({\bm k},z) &=& \frac{-v_z^{(0)}}{z^2 - c_{\perp}^2 k^2 + i z k^2 \Gamma}\ ,
\label{eq:3.10a}\\
\delta T({\bm k},z) &=& \frac{i\delta T^{(0)}}{z + i D_T k^2} 
\nonumber\\
&& + \frac{z (\partial_z T)v_z^{(0)}}{(z + i D_T k^2)(z^2 - c_{\perp}^2 k^2 + i z k^2 \Gamma)}\ ,
\nonumber\\
\label{eq:3.10b}
\eea
\ese
where $z$ is the complex frequency. Transforming back to the time domain, we find
\be
\delta T({\bm k},t) = \delta T^{(0)} e^{-D_T k^2 t} - \frac{(\partial_z T) v_z^{(0)}}{c_{\perp} k}\,\sin(c_{\perp} k t)\,e^{-\Gamma k^2 t/2}.
\label{eq:3.11}
\ee
In the second, non-equilibrium, term we have kept only the leading contribution for $k\to 0$. For the mean-squared displacement, which
can be written
\bse
\label{eqs:3.12}
\be
\langle r^2 \rangle = \frac{-1}{T_0} \left({\bm\nabla}_{\bm k}\right)^2\delta T({\bm k},t)\Bigl\vert_{{\bm k}=0}
\label{eq:3.12a}
\ee
this yields
\be
\langle r^2 \rangle =  6 D_T\,\frac{\delta T^{(0)}}{T_0}\,t + \frac{(\partial_z T)v_z^{(0)}}{2 T_0}\, c_{\perp}^2 t^3\ .
\label{eq:3.12b}
\ee
\ese
This is the result for a solid that is analogous to Eq.~(\ref{eq:3.5}) for a fluid. For the second, non-equilibrium, term only
the leading result is shown, corrections are proportional to $t^2$. The nonequilibrium contribution grows as the time cubed, 
and hence faster than what results from ballistic propagation. This is to be contrasted with the corresponding result in a fluid, 
Eq.~(\ref{eq:3.5}), where the nonequilibrium contribution grows as the time squared.

We finally determine the TTCF in a solid. Performing spatial and temporal Fourier transforms on Eq.~(\ref{eq:3.8b}) yields
\be
u_z({\bm k},\omega) = \frac{-1}{\omega^2 - c_{\perp}^2 k^2 + i\omega \Gamma k^2}\,P_z({\bm k},\omega)
\label{eq:3.13}
\ee
Inserting this in Eq.~(\ref{eq:3.8a}) we have
\be
\delta T({\bm k},\omega) = \frac{1}{\omega + i D_T k^2}\left[ \frac{\omega (\partial_z T) P_z({\bm k},\omega)}{\omega^2 - c_{\perp}^2 k^2 + i\omega\Gamma k^2} + Q({\bm k},\omega)\right]\ .
\label{eq:3.14}
\ee
This yields
\bea
\langle\vert\delta T({\bm k},\omega)\vert^2\rangle &=& \frac{2\kB T_0^2}{c_p}\,\frac{D_T k^2}{\omega^2 + (D_T k^2)^2}
\nonumber\\
&&\hskip -50pt  + \frac{\omega^2 (\partial_z T)^2}{\omega^2 + (D_T k^2)^2}\,
     \frac{2\kB T_0 (\Gamma/\rho){\bm k}_{\perp}^2}{(\omega^2 - c_{\perp}^2 k^2)^2 + \omega^2 \Gamma^2 k^4}\ . \qquad
\label{eq:3.15}
\eea
Integrating over the frequency we finally obtain the solid-state analog to the second term in Eq.~(\ref{eq:1.2}):
\bea
\langle\vert\delta T({\bm k})\vert^2\rangle &=& \int_{-\infty}^{\infty} \frac{d\omega}{2\pi}\,\langle\vert\delta T({\bm k},\omega)\vert^2\rangle
\nonumber\\
&=& \frac{\kB T_0^2}{c_p} + (\partial_z T)^2\,\frac{\kB T_0}{\rho\, c_{\perp}^2 }\,\frac{ {\hat{\bm k}}_{\perp}^2}{k^2}\ .
\label{eq:3.16}
\eea
Comparing with Eq.~(\ref{eq:1.2}), we see that the nonequilibrium effect is similar to that in a fluid, but weaker in the sense that
the TTCF diverges as $1/k^2$ rather than $1/k^4$. For the difference in the equilibrium term ($c_p$ instead of $c_V$), see
Ref.~\onlinecite{pressure_fluctuations_footnote}.

\section{Discussion}
\label{sec:IV}

Consistent with the existence of generic long-ranged correlations in fluids in a NESS, we have shown that there is a novel type of rigidity in the 
macroscopic fluid equations describing perturbations around a NESS. As a consequence of this, the propagation of temperature
perturbations in simple fluids in a temperature gradient is faster than diffusive. In a solid, the corresponding effect is faster than
ballistic.

We conclude with a number of additional remarks:

\begin{enumerate}[leftmargin=*]
\item For an estimate of $v_0$ given by Eq.~(\ref{eq:3.7b}) we use $v_z^{(0)}\approx 5\times 10^4\,$cm/s, a typical thermal velocity, 
$\partial_z T/T_0\approx 0.2\,$cm$^{-1}$, a typical large gradient, and $\nu+D_T\approx 2\times 10^{-2}\,$cm$^2$/s, appropriate for water.
This yields $t_0 \approx 10^{-3}$s and $v_0\approx 10\,$cm/s. This is more than four orders of magnitude smaller than the speed of sound in water, which validates our approximations, which ignored sound waves, {\em a posteriori}. Note, however, 
that for very viscous supercooled liquids,  where $\nu$ is large, $v_0$ can be much bigger.

From Eq.~(\ref{eq:3.5}) we see that the superdiffusive non-equilibrium contribution to $\langle r^2\rangle$ dominates over the 
diffusive equilibrium part for times $t > (24/\pi)(D_T/(\nu + D_T))(\delta T^{(0)}/T_0) t_0$. With $\delta T^{(0)}/T_0 \approx 0.01$,
and parameters again appropriate for water at room temperature, this time scale is on the order of a few $\mu$s. For larger times
the non-equilibrium contribution dominates, and for $t \approx 1$s the root-mean-squared displacement is on the order of a few cm.

For a semi-quantitative estimate of the magnitude of the effect in a solid, we take again $\partial_z T/T_0 \approx 0.2\, \text{cm}^{-1}$,
$v_z^{(0)} \approx 5\times 10^4$ cm/s, and $\delta T^{(0)}/T_0 \approx 0.01$. With $D_T \approx 1\, \text{cm}^2/\text{s}$
and $c_{\perp} \approx 5\times 10^5$cm/s, as appropriate for typical metals, the non-equlibrium term in Eq.~(\ref{eq:3.12b}) dominates
over the diffusive term after a few picoseconds. It is also of interest to compare the former with the speed of a sound wave.
Suppose the perturbing heat pulse is created at the same time and the same location as a sound wave. Then the root-mean-squared
displacement of the heat pulse will overtake the sound wave at a time $t = 2 T_0/\vert(\partial_z T) v_z^{(0)}\vert = t_0/2\pi$, which
is on the order of a millisecond.

We emphasize that the NE effects are very large, leading to correlations on a scale of centimeters and seconds, and these correlations
are generic in the sense that they do not require any fine tuning. By contrast, in order to have the correlation length of an Ising magnet 
reach 1 cm, one must be within roughly $10^{-5}$ of the critical point.

\item The mechanism for producing the anomalous dynamics is very different from the Goldstone mechanism in
an equilibrium system with a broken symmetry. In the latter case, a new soft mode gets created, and often the dynamics of an existing
soft mode are altered, viz., the mode becomes faster due to the rigidity. We recall the simplest case of an observable $\cal{O}$ that is not
conserved and does not couple to any other modes.\cite{Forster_1975} The Kubo function $K$ for that observable then has the structure
\be
K(k,z) = \frac{\chi(k)}{z + i \sigma(k,z)/\chi(k)}
\ee
with $z$ the complex frequency. The quantity $\sigma$ is finite in the limit $k\to 0$, $z\to 0$, since $\cal{O}$ is not conserved. $\chi$ is
the static susceptibility, and if no symmetry is broken, then $\chi(k\to 0)$ is also finite, and there is no soft mode. However, if $\cal{O}$
is a broken-symmetry variable, then $\chi(k\to 0) \propto 1/k^2$ and there is a soft mode with $z \sim k^2$. If ${\cal O}$ were conserved,
then in the absence of a broken symmetry $K$ would have a diffusion pole. Upon breaking the symmetry, an additional soft mode 
would appear, and the existing diffusive mode would change its 
nature. As is obvious from Eqs.~(\ref{eqs:2.4}) and (\ref{eqs:2.5}), this is {\em not} what happens in a NESS. Rather, the nature of 
the existing diffusive mode is unchanged, but the susceptibility that comprises the residue of the diffusion pole becomes long-ranged
as a result of the non-equilibrium fluctuations. This is underscored by the discussion in Sec.~\ref{subsec:II.C}, which shows that the
Langevin force in the fluctuating heat equation gets renormalized, but the dissipative term does not.

\item A long-standing question is whether or not Langevin equations such as Eqs.~(\ref{eqs:2.1}) with fluctuating forces that are
delta-correlated in space, Eqs.~(\ref{eqs:2.2}), can be used to consistently calculate power law correlations of the hydrodynamic
variables in a NESS.\cite{fluctuating_forces_footnote} In Ref.~\onlinecite{Kirkpatrick_Dorfman_2015} it was shown that the effects 
that lead to the long-ranged correlations of the 
hydrodynamic variables do not modify the fluctuating heat and stress currents, so Eqs.~(\ref{eqs:2.1}) and (\ref{eqs:2.2}) can indeed 
be consistently used in a NESS. We note, however, that this conclusion no longer holds if the velocity fluctuations are integrated
out, which makes the fluctuating force in the remaining temperature equation long-ranged, see the discussion after Eqs.~(\ref{eqs:2.12}).

In Sec.~\ref{sec:III} we have effectively shown that the long-ranged behavior in Eqs.~(\ref{eq:1.2}) - (\ref{eqs:1.4}) 
arises from just the deterministic parts of the Langevin equations. That is, by simply solving the averaged equations and
calculating the mean-square displacement, without any reference to fluctuations, one can conclude that the dynamics
are anomalous.

\item Similar long-ranged correlation exist in more complex fluids such as binary mixtures with either a concentration gradient or a thermal 
gradient, \cite{Law_Nieuwoudt_1989, Ortiz_Peluso_Sengers_2004, Ortiz_Sengers_2007} and in wet 
active matter.\cite{Kirkpatrick_Bhattarcharjee_2019} These systems therefore also support the super-diffusive propagation of perturbations.

\item In giving Eqs.~(\ref{eqs:3.4}) we have for simplicity used a continuous Fourier transform rather than a discrete Fourier series in the $z$-direction. 
This simplification places an upper limit on the times for which our explicit results are valid, viz., $t \alt L/v_0$. The super-diffusive non-equilibrium 
contribution in Eq.~(\ref{eq:3.5}) dominates over the diffusive equilibrium contribution for times $t \agt \delta T_0 D_T/T_0v_0^2$. With $L = 10\,$cm,
$v_0 \approx 10\,$cm/s as estimated above, $D_T \approx 0.2\times 10^{-2}\,$cm$^2$/s as appropriate for water, and $\delta T_0/T_0 = 0.03$ 
this yields a large time window $1\,\mu{\text s} \alt t \alt 1\,{\text s}$.

\item We emphasize again that in a fluid there is no propagating mode associated with the spread of a temperature perturbation. Rather,
the temperature gradient couples the temperature fluctuations to the transverse current fluctuations, see Eqs.~(\ref{eqs:2.1}), both of
which are diffusive. However, the coupling results in long-ranged correlations that are reflected in the $1/k^2$ prefactor of the second
term on the right-hand side of Eq.~(\ref{eq:3.2}). This can be seen already at the level of the hydrodynamic equations (\ref{eqs:3.1}):
Since $v_z$ is diffusive, it scales as $1/k^2 \sim z$, and since $\partial_z T$ is constant, this effectively introduces an inhomogeneity
proportional to $1/k^2$ in the diffusion equation for $\delta T$. Upon a Laplace transform, this multiplies the diffusion pole.
As a consequence, a localized temperature perturbation at one point the NE system 
has a measurable effect at a distance that scales with a higher power of time than in the corresponding equilibrium system. The diffusive
dynamics make the divergent prefactor scale as $t$, and therefore the mean-square displacement carries an extra power of $t$ compared 
to the result for the diffusive process. Hence, $\langle r^2\rangle \propto t\times t = t^2$, see Eqs.~(\ref{eq:3.5}) - (\ref{eqs:3.7}).

In a solid, the transverse fluctuations that couple to the temperature fluctuations are propagating, see Eqs.~(\ref{eq:3.8b}), (\ref{eq:3.10a}),
and (\ref{eq:3.13}). Again, the coupling leads to long-ranged correlations that are reflected in the $1/k$ prefactor in the NE term in
Eq.~(\ref{eq:3.11}). The propagating nature of the transverse fluctuations makes this scale as $t$ again, and as a result the mean-squared
displacement scales as $\langle r^2\rangle \propto t \times t^2 = t^3$, Eq.~(\ref{eq:3.12b}). In a solid, the temperature gradient thus has
two distinct effects: First, it couples the constant-pressure temperature (i.e., entropy) fluctuations, which are diffusive in the absence of
the coupling, to a propagating mode. This is somewhat analogous to the coupling between the energy density and the superfluid velocity
that creates the second-sound mode in a superfluid. Second, it leads to long-ranged correlations that make the temperature/entropy
fluctuations `supersonic' in the sense that $\langle r^2\rangle \propto t^3$ rather than $t^2$.

In the calculation of the TTCF the divergent prefactor of the coupled mode effectively gets squared, and hence the TTCF diverges as $1/k^4$ 
in a fluid, Eq.~(\ref{eq:1.2}), and as $1/k^2$ in a solid, Eq.~(\ref{eq:3.16}).

\end{enumerate}


\begin{thebibliography}{30}
\bibitem[{tra()}]{transverse_footnote}
\bibinfo{note}{That is, $v_z$ is the $z$-component of ${\bm v}_{\perp}$, which
  is the part of the fluctuating velocity ${\bm v}({\bm k},t)$ that is
  perpendicular to the wavenumber ${\bm k}$. Equation~(\ref{eq:2.1b}) can be
  obtained from the general Langevin equation for $v_{\beta}({\bm k},t)$, see
  Ref.~\onlinecite{Landau_Lifshitz_VI_1987}, by multiplying it by the projector
  ${\mathcal P}_{\alpha\beta} = \delta_{\alpha\beta} - {\hat k}_{\alpha}{\hat
  k}_{\beta}$, where $\hat{\bm k} = {\bm k}/k$, and then setting $\alpha = z$.}
\end{thebibliography}



\end{document}